\documentstyle[12pt]{article}

\def\be{\begin{equation}}
\def\ee{\end{equation}}
\def\ba{\begin{array}{c}}
\def\baa{\begin{array}{ll}}
\def\ea{\end{array}}
\def\p{\partial}

\def\ben{$$}
\def\een{$$}
 
\begin{document}

\titlepage

  \begin{center}{\Large \bf

Solvable ${\cal PT}-$symmetric model with a tunable interspersion
of non-merging levels.

 }\end{center}

\vspace{5mm}

  \begin{center}

Miloslav Znojil\footnote{ e-mail: znojil@ujf.cas.cz}

 \vspace{3mm}

\'{U}stav jadern\'e fyziky AV \v{C}R, 250 68 \v{R}e\v{z}, Czech
Republic\\

 \vspace{3mm}

\end{center}

\vspace{5mm}

\section*{Abstract}

We study the spectrum in such a ${\cal PT}-$symmetric square well
(of a diameter $L \leq \infty$) where the ``strength of the
non-Hermiticity" is controlled by the two parameters, viz., by an
imaginary coupling $ig$ and by the distance $\ell <L$ of its onset
from the origin. We solve this problem and confirm that the
spectrum is discrete and real in a non-empty interval of $g \leq
g_0(\ell,L)$. Surprisingly, a specific distinction between the
bound states is found in their asymptotic stability/instability
with respect to an unlimited growth of $g$ beyond $g_0(\ell,L)$.
In our model, {\em all} of the low-lying levels remain
asymptotically unstable at the small $\ell \ll L$ and finite $L$
while only the stable levels survive near $\ell \approx L< \infty\
$ {\em or } in the purely imaginary force limit with $0<\ell<L=
\infty$. In between these two extremes, an unusual and tunable,
variable pattern of the interspersed ``robust" and ``fragile"
subspectra of the real levels is obtained.

 \vspace{9mm}

\noindent PACS  03.65.Ge, 03.65.Ca, 02.30.Tb, 02.30.Hq

\noindent MSC 2000: 81Q05, 81Q10, 46C20, 47B50, 34L40



 \newpage

\section{Introduction}

Around 1992, Daniel Bessis succeeded in attracting attention of a
few people to a certain toy Hamiltonian (with some relevance in
quantum field theory) which appeared to produce the real and
discrete spectrum of energies in spite of being manifestly
non-Hermitian \cite{DB}. A few years later, Bender and Boettcher
returned to his mind-boggling problem and published a numerical
study \cite{BB} of the whole class of the perceivably more general
one-dimensional Schr\"{o}dinger equations
 \be
 \left [ -\frac{d^2}{dx^2} + V(x) + i\,W(x)
 \right ]\,
 \psi(x)=E\,\psi(x)
 \label{basic}
 \ee
where, in our present perspective, the real component of the
potential was assumed spatially symmetric while its
Hermiticity-violating partner was chosen as spatially
antisymmetric,
 \ben
  {\cal P} V(x) {\cal P} = V(-x) = +V(x), \ \ \ \ \
 {\cal P} W(x) {\cal P} = W(-x) = -W(x).
 \een
The latter study confirmed that the similar models [exhibiting,
obviously, the parity (${\cal P}$) times time-reversal (${\cal
T}$) symmetry] may possess {\em both} the purely real and
partially (or, perhaps, completely) complex spectra. The Bender's
and Boettcher's Figure~1 ({\em loc.~cit.}) illustrated the
existence of the spectrum which proved ``robustly real", i.e.,
real in a wide range of parameters of their ``massless" ${\cal
PT}-$symmetric model. In contrast, a merely slightly modified
``massive" ${\cal PT}-$symmetric model of their Figure~3 ({\em
loc.~cit.}) behaved quite differently. The values of many of its
energy levels proved extremely sensitive to the very small
variations of the parameters and, moreover, even the very reality
of some energies proved ``fragile" in the sense that after a very
small change of a parameter of the model, certain energy pairs
merged and disappeared forming, presumably, the complex conjugate
pairs. At present, many more similar and more or less purely
numerical examples exists (cf., e.g., the recent paper
\cite{Frank} for a sample of references).

The recent progress in our understanding of the various ${\cal
PT}-$symmetric quantum Hamiltonians $H$ may be briefly summarized
as an observation that their symmetry is important. Firstly, it
was established that the time-reversal-type antilinear operator
factor ${\cal T}$ merely mediates the Hermitian conjugation $A \to
A^\dagger$ \cite{AM,BBJerr}. The role of parity ${\cal P}$ is more
subtle and seems to offer the main mathematical key to the study
of the ${\cal PT}-$symmetric quantum Hamiltonians $H$ within the
so called Krein-space theory (cf., e.g., ref. \cite{Langer} for a
nice as well as concise introduction to this language).

On the background of these {\em mathematical} observations, the
formalism lost its originally highly enigmatic features in the
context of {\em physics}. During the last two or three years, the
use of the ${\cal PT}-$symmetric quantum models has in fact been
accepted as just opening new horizons within the standard Quantum
Mechanics. At present, virtually all the people active in the
field would agree that it is only necessary to make the resulting
physical picture complete by a revitalization of its probabilistic
contents and tractability. This is being achieved via an
introduction of the ``missing" (and, in fact, quite nontrivial)
metric $\eta \neq I$ in the Hilbert space of states
\cite{Langer,MB,AmerJ}.

The temporary doubts and puzzles related, typically, to the
applicability of the formalism look, at least roughly, clarified.
One feels urged to return to many recently neglected and
apparently evasive and mathematically more subtle questions like
the problems of the robustness/fragility of the individual
energies or of a global typology of the spectra. We believe that
it is time for their deeper and more technical study via, say,
simplified and, first of all,  {\em non-numerically tractable}
models. A new one, with rather surprising properties and
descriptive features of the spectrum, is to be proposed and
analyzed in what follows.

\subsection{Non-Hermitian square-well-type models }

Within ${\cal PT}-$symmetric Quantum Mechanics a one-parametric
non-Hermitian square well (NSW) model has been described in
ref.~\cite{sqw}. A key merit of the NSW model lies in a
combination of its straightforward mathematical solvability with
an exceptional transparency of its applications. In this way, the
NSW model was able to offer an insight into the mechanism of the
spontaneous ${\cal PT}-$symmetry breaking \cite{Geza}. Next, due
to its elementary character, the NSW model has been selected by
Bagchi et al \cite{Quesne} as a starting point of a systematic
supersymmetric generation of solvable non-Hermitian Hamiltonians
with ${\cal PT}-$symmetry and real spectra. Last but not least,
Mostafazadeh and Batal \cite{MB} choose the NSW model in their
very recent illustrative application of the ${\cal PT}-$symmetric
Quantum Mechanics in its present, mathematically as well as
physically  more or less consistent updated form (readers may
consult some of the available reviews for more details
\cite{pseudo}).

In our recent paper \cite{fragil} we revealed that a certain
``hidden" shortcoming of the NSW model may be seen in its
``fragility", i.e., in an instability of all the higher energy
levels with respect to a certain highly speculative form of a
complex-coordinate perturbation. Although such an observation does
not have any {\em immediate} impact on the applications of the NSW
model in refs. \cite{MB,Geza,Quesne}, certain doubts survive
concerning the possible manifestations of some more serious
instabilities in some of the generalized, NSW-type (NSWT) models.

For our present purposes let us vaguely characterize the latter
NSWT potentials as piecewise constant. Then we may immediately
recollect the existence of several ``user-friendly" NSWT examples
incorporating square-well models on a compact domain \cite{JakubZ}
or systems based on the use of point interactions \cite{FeiZ}.
Unfortunately, even within this class, the expectations concerning
the stability of the spectrum are not always fulfilled. One may
recollect, e.g., a spontaneous complexification of the high-lying
part of many NSWT spectra as detected in very early numerical
studies of certain particular potentials in ref.~\cite{BBjmp}. The
phenomenon looks puzzling and makes all the NSWT models worth a
more detailed non-numerical study.

\subsection{The choice of a specific example }


In applied quantum mechanics the construction of the majority of
phenomenological models relies quite heavily on the correspondence
principle which tries to connect each quantum model with its
classical predecessor. ${\cal PT}-$symmetric Quantum Mechanics
offers a weakening of this connection \cite{experimental}. The
operator of parity ${\cal P}$ is indefinite so that, as we already
mentioned, the formalism requires an explicit {\em additional}
construction of a Hamiltonian-dependent {positively definite}
metric $\eta>0$ in Hilbert space. Equivalently, this may be
mediated by the construction of a quasi-parity ${\cal Q}$
\cite{ptho} or charge ${\cal C}$ \cite{BBJ}, both defined as a
product ${\eta}{\cal P}$. In practical calculations this means
that the metric is often being introduced in a suitably factorized
form~\cite{Cannata}.

It is worth adding that the quasi-parity in $\eta = {\cal QP}$ is
easily defined in some exactly solvable examples \cite{ptho} while
the charge in $\eta = {\cal CP}$ has immediate connotations in
field theory \cite{BBJ}. In between these two extremes the authors
of ref. \cite{MB} revealed that the application of the formalism
to the particular NSW model proves facilitated by a perturbative
connection between the NSW model and a Hermitian square well.
Their construction of ${\eta}^{(NSW)}$ profited from the existence
of a finite-dimensional matrix approximation of the non-Hermitian
part of the NSW Hamiltonian. A transition to the extended NSWT
class of models looks promising and co-motivates also our present
project.

Within such a framework we intend to pay attention to the family
of  Schr\"{o}dinger equations (\ref{basic}) where the interaction
is non-Hermitian but manifestly ${\cal PT}-$symmetric. For the
sake of definiteness we shall contemplate the less interesting
real part of the potential just in the most elementary infinitely
deep square-well form,
 \be
  V(x) = \left \{  \begin{array}{l} + \infty\\ \ \  0 \\
 + \infty \ea \right . \ \ \ \ \ \ \ {\rm for\ \  \ } \ \ \
 \left \{
  \begin{array}{l}
 \ x\
 > L \\
  -L < x
 < L \\
 \ x\
 < -L\,.
   \ea
   \right .
 \label{SQWr}
  \ee
This means that all our wave functions have to vanish at its
walls,
 \be
 \psi(-L) = \psi(L) = 0\,.
 \ee
By adding any imaginary interaction we break the Hermiticity of
the Hamiltonian. By doing so in the ${\cal PT}-$symmetric manner
we preserve a chance and good hope of having the energies real
\cite{BB}.

For the sake of definitness and in a way generalizing the NSW
model of ref. \cite{sqw} we shall assume that the
Hermiticity-breaking term $W$ is composed of two purely imaginary
steps which both vanish inside a subinterval $(-\ell, \ell)$ of
the interval $(-L,L)$,
 \be
  W(x) = \left \{  \begin{array}{l} + ig,\\
  \ \ \ 0
  \\ -
 ig \ea \right . \ \ \ \ \ \ \ {\rm for\ \  \ } \ \ \
 \left \{
  \begin{array}{l}
 {\rm Re}\ x\
 > \ell >0\,,\\
{\rm Re}\ x\
 \in (- \ell,\ell)\,, \\
 {\rm Re}\ x\
 < - \ell\,.
   \ea
   \right .
 \label{SQWe}
  \ee
{\it A priori}, the strength of the Hermiticity-violating
imaginary force may be expected proportional to the coupling $g>0$
and inversely proportional to $\ell < L$.

Our interest in the particular two-parametric model (\ref{SQWe})
results from the obvious need of an enhancement of flexibility of
its one-parametric NSW predecessor and also from the lasting
possibility of its rigorous mathematical description by means of
the efficient moving-lattice method of ref.~\cite{fragil}
(reviewed also briefly in Appendix A below). Among additional
purposes of the study of the similar NSWT models one may list a
search for reliable comparisons between different potentials
revealing, hopefully, some new, unnoticed characteristic features
of their spectra. One would like to understand, i.a., how the
details of the shape of $W(x)$ could influence the stability of
the spectrum, or how one could control the domain of parameters
where all the energies remain real.

Some of the NSWT studies have been motivated by their potential
capacity of mimicking the properties of unsolvable models and, in
particular, of one of the most popular ${\cal PT}-$symmetric toy
interactions $W(x) = ix^3$ \cite{Caliceti}. Some parallels are
definitely there since in the latter unsolvable case the spectrum
was proved real, non-negative and discrete \cite{DDT}. Of course,
there are always good reasons for an introduction of more
parameters in NSW. Thus, the new freedom of a weakening of the
non-Hermiticity by the choice of $\ell > 0$ might simulate
analogies with the Bender's and Boettcher's generalized ${\cal
PT}-$symmetric family $W(x) = -(ix)^{3-\mu}$ characterized by an
abrupt change of its spectral properties at $\mu=1$ and by the
spontaneous complexification of all the sufficiently high-lying
energies inside the interval $\mu \in (1,2)$ of the
shape-parameter~\cite{BB}.

The possibility of the latter correspondence passes an easy test
at $\ell = 0$ and $L = \infty$ when the general solutions of our
Schr\"{o}dinger eq. (\ref{basic}) are mere exponentials at any
real $g> 0$. Once we demand that they vanish in infinity we have
  \be
 \psi(x) =
 \left \{
 \begin{array}{llll}
 B_+\,\exp (-\sigma\,x)
  ,&\sigma^2=ig-E\,,&
{\cal R}e \sigma> 0,& x \in (0,\infty),\\
 B_-\,\exp (\sigma'\,x)
  ,&
\sigma'^2=-ig-E\,,&{\cal R}e \sigma'> 0,& x \in (-\infty,0).
   \ea
   \right .
 \label{ansareah}
  \ee
When $x \to 0^\pm$ the coincidence of the right and left limit of
$\psi(x)$ itself specifies the normalization, $B_+=B_-$, while the
second matching rule $\psi'(0^+)=\psi'(0^-)$ implies that $\sigma
= - \sigma'$, i.e.,  equation (\ref{ansareah}) has no solutions at
$g>0$. It is of no avail to admit that ${\cal R}e \ \sigma \to 0$
and $ {\cal R}e \ \sigma' \to 0$ and to employ the scattering
boundary conditions since, unless $g=0$, the matching-compatible
states remain always incompatible with our differential
Schr\"{o}dinger equation on a half-line.

We may conclude that both the discrete and continuous spectra are
empty at $\ell=0$ for $g>0$ and $L=\infty$. This re-confirms our
above expectations since the emptiness of the spectrum also
characterizes the Bender's and Boettcher's toy interaction $W(x) =
-(ix)^{3-\mu}$ at the Herbst's extreme shape parameter $\mu=2$
\cite{Herbst}. At the same time, the spectrum abruptly ceases to
be empty at $\mu < 2$ \cite{BB} as well as at $\ell > 0$ while $L
= \infty\ $ (cf. the proof of this assertion as given in Appendix
B below).

\section{The method
\label{sectiontri}}


\subsection{Wave functions and their matching}

As long as our potential is piecewise constant at $0 < \ell < L<
\infty$ we may postulate
 \be
 \psi(x) = \left \{
 \baa
  \psi_-(x) = B_-\,\sinh \kappa^*(L+x),\ \
  & \ \ x \in (-L,-\ell),\\
  \psi_0(x) = {C} \,\cos k\,x + i\,{D}\,\sin k\,x,\ \
  & \ \ x \in (-\ell,\ell),\\
  \psi_+(x) = B_+\,\sinh \kappa\,(L-x),\ \
  & \ \ x \in (\ell,L)\,
 \ea
 \right .
 \label{ansatz}
 \ee
where $\kappa = s + it, \  E = k^2 = t^2-s^2, \ g = 2st > 0$ and
where $s$, $t$ and $k$ are assumed real and, for the sake of
definiteness, positive. In the other words, we assume that within
a not yet specified non-empty domain of parameters $g$ and $\ell$
the ${\cal PT}$ symmetry of the wave functions remains unbroken.
In the way proposed in ref. \cite{sqw} we prescribe the phase,
 \ben
 \psi(x) = real\ symmetric + imaginary\ antisymmetric\,
 \een
and deduce that ${C}$ and ${D}$ are real. Next, we differentiate
 \ben
 \psi'(x) = \left \{
 \baa
  \psi_-'(x) =\kappa^*\, B_-\,\cosh \kappa^*(L+x),\ \
  & \ \ x \in (-L,-\ell),\\
  \psi_0'(x) = -k\,{C} \,\sin k\,x + i\,k\,{D}\,\cos k\,x,\ \
  & \ \ x \in (-\ell,\ell),\\
  \psi_+'(x) = -\kappa\,B_+\,\cosh \kappa\,(L-x),\ \
  & \ \ x \in (\ell,L)\,,
 \ea
 \right .
 \label{ansatzpr}
 \een
and write down the following four matching conditions,
 \ben
   \baa
  \psi_-(-\ell) = \psi_0(-\ell), \ \ i.e.,\ \ &\ \
   B_-\,\sinh \kappa^*(L-\ell)=
    {C} \,\cos k\,\ell - i\,{D}\,\sin k\,\ell,
  \\
  \psi_-'(-\ell) = \psi_0'(-\ell), \ \ i.e.,\ \ &\ \
     \kappa^*\, B_-\,\cosh \kappa^*(L-\ell)=
   k\,{C} \,\sin k\,\ell + i\,k\,{D}\,\cos k\,\ell,\\
  \psi_+(\ell) = \psi_0(\ell), \ \ \ \ \ \ \,  i.e.,\ \ &\ \
   B_+\,\sinh \kappa\,(L-\ell)= {C} \,\cos k\,\ell + i\,{D}\,\sin
   k\,\ell,\\
  \psi_+'(\ell) = \psi_0'(\ell), \ \ \ \ \ \ \,  i.e.,\ \ &\ \
   -\kappa\,B_+\,\cosh \kappa\,(L-\ell)=
     -k\,{C} \,\sin k\,\ell + i\,k\,{D}\,\cos k\,\ell\,.
 \ea
 \label{ansatzma}
 \een
Two of them define the (complex) values of $B_\pm$ so that we are
left with the pair of the matching constraints,
 \ben
 \left ( k\,{C} \,\sin k\,\ell + i\,k\,{D}\,\cos k\,\ell \right )
 \sinh \kappa^*(L-\ell)
 =
  \left ( {C} \,\cos k\,\ell - i\,{D}\,\sin k\,\ell \right )
 \kappa^*\, \cosh \kappa^*(L-\ell)
 \label{druhar}
 \een
 \ben
 \left ( k\,{C} \,\sin k\,\ell - i\,k\,{D}\,\cos k\,\ell
 \right )
 \sinh \kappa\,(L-\ell)
=
 \left ({C} \,\cos k\,\ell + i\,{D}\,\sin k\,\ell \right )
 \kappa\,\cosh \kappa\,(L-\ell)\,.
 \label{jednar}
 \een
These two relations are complex conjugate of each other so that we
have to consider just one of them, say,
 \ben
 \left ( k\,{C} \,\sin k\,\ell - i\,k\,{D}\,\cos k\,\ell
 \right )
 \sinh  (s + it)\,(L-\ell)
 =
 \een
 \be
 =
 \left ({C} \,\cos k\,\ell + i\,{D}\,\sin k\,\ell \right )
 \kappa\,\cosh (s + it)\,(L-\ell)\,.
 \label{jedarde}
 \ee
with $k \geq 0$.

\subsection{Matching equations in the
$\sigma - \tau - \varrho$ space}

After we abbreviate $\sigma =s \,(L-\ell)$, $\tau =t \,(L-\ell)$
and $\varrho =k\,\ell$, equation (\ref{jedarde}) reads
 \ben
 \varrho
  \,(L-\ell)\,
 \left ( {C} \,\sin \varrho - i\,{D}\,\cos \varrho
 \right )
 \left [
 \sinh  \sigma
 \cos \tau
+ i\,
 \cosh \sigma
 \sin \tau
 \right ]
= \een
 \be
 =
 \ell\,
 (\sigma+i\,\tau)\,
 \left ({C} \,\cos \varrho + i\,{D}\,\sin \varrho \right )
 \,
  \left [
 \cosh  \sigma
 \cos \tau
+ i\,
 \sinh \sigma
 \sin \tau
 \right ]
 \,.
 \label{jerderoz}
 \ee
We have to keep in mind that
 \ben
 \tau^2 = \sigma^2 + \frac{(L-\ell)^2}{\ell^2}\varrho^2
 \,
 \een
while the respective real and imaginary parts of eq.
(\ref{jerderoz}) have to be treated as independent equations
 \ben
 \varrho
  \,(L-\ell)\,
 \left ( {C} \,\sin \varrho
  \sinh  \sigma
 \cos \tau
+ {D}\,\cos \varrho \,
 \cosh \sigma
 \sin \tau
 \right )
=
\een \ben
=
 \ell\,
 \left [
 \sigma\,
 \left (
 {C} \,\cos \varrho
 \,
   \cosh  \sigma
 \cos \tau
-{D}\,\sin \varrho \,
 \sinh \sigma
 \sin \tau
\right ) \right . - \een \be
 - \left .
 \tau\,
  \left (
 {C} \,\cos \varrho
 \,
 \sinh \sigma
 \sin \tau
 +
 {D}\,\sin \varrho
 \,
 \cosh  \sigma
 \cos \tau
 \right )
 \right ]
\,
 \label{jerozdat}
 \ee
and
 \ben
 \varrho
  \,(L-\ell)\,
 \left ( {C} \,\sin \varrho
 \cosh \sigma
 \sin \tau
  -{D}\,\cos \varrho \,
 \sinh  \sigma
 \cos \tau
 \right )
= \een
 \ben
 =
 \ell\,
\left [
 \sigma\,
 \left (
 {C} \,\cos \varrho
 \,
 \sinh \sigma
 \sin \tau
 +
 {D}\,\sin \varrho
 \,
 \cosh  \sigma
 \cos \tau
 \right )
 \right .
+ \een
 \be
 +
 \left .
 \tau\,
\left (
 {C} \,\cos \varrho
 \,
   \cosh  \sigma
 \cos \tau
-{D}\,\sin \varrho \,
 \sinh \sigma
 \sin \tau
\right )
 \right ]
 \,.
 \label{jerozbit}
 \ee
In the next step we notice that the latter equations form a linear
algebraic homogeneous set for the two coefficients ${C}$ and
${D}$. They possess a nontrivial solution if and only if the
secular determinant ${\cal D}$ vanishes. After we abbreviate
${\Omega}= \tan \varrho$ (= a quickly oscillating function of
$\varrho$), $T= \tan \tau$ (= a quickly oscillating function of
$\tau$) and $\Sigma= \tanh \sigma$ (= a monotonous and bounded
function of $\sigma$) we can evaluate ${\cal D}$. After a lengthy
calculation the secular condition ${\cal D}=0$ acquires the
following compact form
 \be
 X(\sigma) + Y(\tau) + F(R) \,[x(\sigma) + y(\tau)]=0
 \label{sehen}
 \ee
where
 \ben
 X(\sigma)=  \frac{1+\Sigma^2}{1-\Sigma^2}\,\sigma^2
 = \sigma^2\,\cosh 2\,\sigma\,,
 \een
 \ben
 Y(\tau)=  \frac{1-T^2}{1+T^2}\,\tau^2=
  \tau^2\,\cos 2\, \tau\,,
 \een
 \ben
 x(\sigma)=  \frac{\Sigma}{1-\Sigma^2}\,\sigma
 =  \frac{1}{2}\,\sigma\,\sinh 2\,\sigma\,,
 \een
 \ben
 y(\tau)=  \frac{T}{1+T^2}\,\tau=  \frac{1}{2}\,\tau\,\sin
 2\,\tau\,,
 \een
 \ben
 F(R)=  \frac{1-{\Omega}^2}{{\Omega}}\,R\,=  \frac{2\,R}{\tan 2\,\varrho}\,,
 \ \ \ \ \ \varrho=\varrho(R)=\frac{\ell}{L-\ell}\,R\,.
 \een
We may re-scale our coupling $\ g=2\,Z/(\L-\ell)^2$ and conclude
that our $Z-$independent secular equation (\ref{sehen}),
 \be
  {\sin 2\,\varrho(R)}
  \,\left [
 \sigma^2\,\cosh 2\,\sigma + \tau^2\,\cos 2\, \tau \right ]
  +R\,  {\cos 2\,\varrho(R)}
  \,\left [
  \sigma\,\sinh 2\,\sigma
   +  \tau\,\sin 2\,\tau
   \right ]=0\,
   \label{secularia}
 \ee
only has to be complemented by the two trivial constraints
 \be
 \sigma\,\tau=Z\,,\ \ \ \ \ \ \ \ \ \  \
 \tau^2-\sigma^2=R^2\,.
 \label{constraintsdve}
 \ee
The triplets of roots $R_n$, $ \sigma_n$ and $\tau_n$ of this
triplet of equations with $ n = 0, 1, \ldots$ define all the
bound-state energies $E_n$ by the elementary formula
 \be
 E_n = \frac{1}{(L-\ell)^2}\,
 R^2_n \,\equiv \, \frac{1}{(L-\ell)^2}\,\left (
 \tau^2_n-\sigma^2_n
 \right ) \ .
 \label{eneriad}
 \ee
In an indirect check of the recipe we may recollect its $\ell \to
0$ (i.e., $\varrho \to 0$) limit and conclude that our present eq.
(\ref{secularia}) degenerates smoothly and correctly back to the
known secular $\ell=0$ equation \{cf. eq. Nr. (9) in ref.
\cite{sqw}\}.

\subsection{Matching in the
moving-lattice representation \label{pohyb}}

The basic tool for a rigorous analysis of the form of the
solutions of our matching constraints is the moving-lattice method
of ref. \cite{fragil} as reviewed in Appendix A below. Skipping
the majority of details let us only note that for an analysis of
this type, one of the recommended techniques seems to be the
reduction of the problem to $\sigma - \tau$ plane. Preserving the
definition of $\tau=\tau(N,t)$ of Appendix A and replacing the
definition of $\sigma=\sigma(N,t)$ by another formula,
 \ben
 \sigma
 =\sigma(N,t,K,r)
  =\pi\,\times\,
\sqrt{[{  N + t}]^2 + \left [ \frac{L-\ell}{2\,\ell}\,(K+r) \right
]^2
 }\ ,
 \een
we eliminate the coordinate $R$. A shortcoming of this approach is
that our matching condition (\ref{secularia}) transferred into the
$\sigma - \tau$ plane has to be understood as the following
quadratic equation for $\tau$,
 \be
 \Phi_t\,\tau^2+\omega_{K,r,t}\,\tau + \Omega_{K,r,t}(\sigma)=0
 \label{secudd}
  \ee
where we abbreviated
 \ben
  \omega_{K,r,t} =
  \frac{(L-\ell)\,\pi\,\Psi_t}{2\ell\,\Xi_r}\,(K+r),\ \ \ \
     \Omega_{K,r,t}(\sigma)=
 {\Xi}_r\,
 \left [
 \sigma^2\,\cosh 2\,\sigma +
 \frac{ \omega_{K,r,t} }{\Psi_t}\,
 \sigma\,\sinh 2\,\sigma
 \right ]
   \,.
 \een
This defines $\tau=\tau_{K,r,t}(N)$ on the lattice, the ``motion"
of which will be controlled not only by $t$ and $r$ but also, not
so strongly, by $K$. Technically, the price to be paid is still
reasonable - we get the closed form of the matching-compatible
function $\tau=\tau(\sigma)$ as the two well known root formulae
from eq. (\ref{secudd}). Nevertheless, significant simplifications
of the resulting picture may be mediated by the direct inspection
of the equations in question.

\section{Solutions \label{tristar}}

\subsection{Matching equations in the $\sigma
- \tau$ plane \label{tristarty}}

Building far-reaching analogies with the $\ell=0$ special case
would be misleading because the form of our matching constraint
(\ref{secularia}) is discontinuous in the limit $\ell \to 0$.
Thus, let us assume that $\ell \neq 0$ and study eq.
(\ref{secularia}) in its full-fledged form. Firstly, we abbreviate
${\cal M}(\sigma,\tau)=\sigma\,\sinh 2\,\sigma + \tau\,\sin
2\,\tau$ and ${\cal N}(\sigma,\tau)= \sigma^2\,\cosh 2\,\sigma +
\tau^2\,\cos 2\, \tau$ and re-write our matching constraint
(\ref{secularia}) as the secular equation
 \be
 {\cal D}(\sigma,\tau,R) =Q(\sigma,\tau)
  + \frac{\tan 2\,\varrho(R)}{R}
  =0, \ \ \ \ \ \ \ \
  \
  Q(\sigma,\tau)=
   \frac{{\cal M}(\sigma,\tau)}{{\cal N}(\sigma,\tau)}
   \,.
   \label{seculibed}
 \ee
This enables us to formulate several obvious observations.

\begin{itemize}

 \item[[O1${\rm ]}$]
  The shape of both the functions
${\cal M}(\sigma,\tau)$ and ${\cal N}(\sigma,\tau)$ of two
variables is easily deduced using their separability, ${\cal
X}(\sigma,\tau)={\cal X}(\sigma,0)+ {\cal X}(0,\tau)$, ${\cal X} =
{\cal M,N}$.

 \item[[O2${\rm ]}$]
  The smoothness of the $\sigma-$ and $\tau-$dependence of the
denominator ${\cal N}(\sigma,\tau)$ facilitates also the
determination of the shape of ${\cal F}(\sigma,\tau)=1/{\cal
N}(\sigma,\tau)$.

 \item[[O3${\rm ]}$]
 In $\sigma - \tau$ plane we may visualize the shape of the
second fraction in (\ref{seculibed}) as a function which is
constant along hyperbolas $R(\sigma,\tau)=
\sqrt{\tau^2-\sigma^2}=$ {\it fixed}.

\end{itemize}

 \noindent
All these innocent-looking observations have several far-reaching
though not always obvious consequences and form in fact a
background for a rigorous analysis of the spectrum.

\subsection{A rigorous graphical interpretation of
  $Q(\sigma,\tau)$}

In more detail, observation [O1] means that the surfaces defined
by the two non-negative function(s) ${\cal X}(\sigma,0) \geq 0$
have the form of the two only slightly different parabolic valleys
with the same degenerate minimum (= zero) which coincides with the
axis $\sigma=0$. The pertaining second components ${\cal
X}(0,\tau)$ differ more from each other but both are adding a
structurally similar perpendicular set of infinitely many parallel
hills and valleys possessing a steadily increasing (though always
finite) amplitude. As an obvious result of the superposition, both
the resulting surfaces ${\cal X}(\sigma,\tau)$ cross the zero
plane merely along certain ovals $O_n^{{\cal X}}$, and both of
them only get negative in their interior.

The precise shape of these ovals (numbered by $n=0, 1, \ldots$)
may fully rigorously be determined using the moving-lattice method
(cf. Appendix A) but even without any use of the moving lattices
the qualitative character of their shape is obvious and we may
conclude that the zero lines of ${\cal M}(\sigma,\tau)$ and ${\cal
N}(\sigma,\tau)$ form the families of ovals $O_n^{{\cal M}}$ and
$O_n^{{\cal N}}$ located within the stripes of $\tau \in
[(n+1/2)\pi, (n+1)\pi]$ and $\tau \in [(n+1/4)\pi, (n+3/4)\pi]$,
respectively. All of them are symmetric with respect to the
reflection $\sigma \to -\sigma$ and their size in the $\sigma$
direction increases with $\tau$.

Examples of these structures may be found in both refs. \cite{sqw}
and \cite{fragil} and another illustration appears in Figure~1
here. In fact, the Figure displays another surface $Q(\sigma,\tau)
={\cal M}(\sigma,\tau)/{\cal N}(\sigma,\tau)$ (within a narrow
window of $0 \leq Q \leq 0.05$) but the shape of the curve where
${\cal M}$ {\bf v}anishes ($O_1^{{\cal M}}\equiv V_1$) appears
there clearly since the denominator ${\cal F}(\sigma,\tau)=1/{\cal
N}(\sigma,\tau)$ has its zeros, generically, elsewhere (cf.
observation [O2]). Besides the oval $V_1$ (and a part of
$O_0^{{\cal M}}\equiv V_0$) the picture displays another oval
$O_1^{{\cal N}}\equiv D_1$ of the zeros of the {\bf d}enominator
${\cal N}$. Incidentally it lies within the chosen interval of
$\tau \in (3,7)$ and remains visible due to a numerical artifact
of a spurious projection of an infinite {d}iscontinuity of the
function ${\cal F}(\sigma,\tau)$.

Although the visibility of the discontinuities reflects just an
imperfection of the graphical representation of the surface, in
will prove useful in what follows.

\subsection{The role of the second component of ${\cal
D}[\sigma,\tau,R(\sigma,\tau)]$ \label{tangensy}}

The presence of the subsurface generated by the second,
$R-$dependent component ${\cal D}_{(R)}[R(\sigma,\tau)]$ in
eq.~(\ref{seculibed}) does not violate the separation between
$\sigma$ and $\tau$ too much (cf. observation [O3]). At the
smallest absolute values of $\sigma$ we may safely return to the
approximation of ${\cal D}_{(R)}[R(\sigma,\tau)]$ by a function of
a single variable, $[\tan 2\,\varrho(R)]/{R} \approx [\tan
2\,\ell\,\tau/(L-\ell)]/{\tau} $. This picture only becomes
deformed, at the larger $\sigma$, by being bent to the right,
i.e., along hyperbolas $R(\sigma,\tau) =constant$.

A clear understanding of the $\tau-$dependence of the whole
surface ${\cal D}[\sigma,\tau,R(\sigma,\tau)]$ will be obtained
when we distinguish between the domain of the ``small $\tau$"
\{where $[\tan 2\,\ell\,\tau/(L-\ell)]/{\tau} \approx
2\,\ell/(L-\ell)$ is positive and virtually constant\}, ``medium
$\tau$" \{with the repeated quick growth of the curve $[\tan
2\,\ell\,\tau/(L-\ell)]/{\tau} $ from minus infinity up to plus
infinity within each interval of the constant length $\triangle
\tau = \pi\,(L-\ell)/2\,\ell$\} and ``large $\tau$" \{where the
values of  ${\cal D}_{(R)}[R(\sigma,\tau)] \approx 1/\tau$ become
very small up to the very thin layers near the singularity
hyperbolas $H_n$\}. Due to the local dominance of the latter
singularities $H_n$ at any $n = 0, 1, \ldots$ it is easy to
imagine that the sign of the whole function ${\cal
D}[\sigma,\tau,R(\sigma,\tau)]$ is positive and negative in their
left and right vicinity, respectively. This ``rule of thumb"
enables us to deduce the sign of the whole function ${\cal
D}[\sigma,\tau,R(\sigma,\tau)]$ in all our Figures.


\subsection{The left-moving hyperbolic discontinuities $H_n$  }

In the domain of the small shifts $\ell \ll 1$ the numerical
values of the $R-$dependent component ${\cal
D}_{(R)}[R(\sigma,\tau)]$ of eq. (\ref{seculibed}) remain almost
constant and small. In this regime the above-mentioned
``small-$\tau$" constraint $\tau \ll (L-\ell)/\ell\ $ is not
particularly restrictive so that the matching-compatible roots of
equation ${\cal D}=0$ remain very similar to their $\ell=0$
predecessors in quite a large leftmost portion of the $\sigma -
\tau$ plane. In our notation, the first few ovals $O_n^{{\cal D}}
\equiv V_n$ of the zeros of the secular determinant stay only
perturbatively shifted and deformed by an increase of $\ell\ll 1$.

With the growth of $\ell$ or $\lambda=\ell/(L-\ell)$ the leftmost
discontinuity-hyperbola $H_0$ of the surface ${\cal
D}[\sigma,\tau,R(\sigma,\tau)]$ moves to the left and emerges in
the right half of Figure~2 where we choose the scale-independent
parameter $\lambda = 11/40$ which corresponds to $\ell=11\,L/51$.
This means that we are just leaving the domain of the small shifts
$\ell \ll 1$ so that the deformation of the nodal oval $O_1^{{\cal
D}}\equiv V_1$ becomes perceivable, caused by the closeness of
$H_0$ to the  $\ell-$independent discontinuity oval $D_1 \equiv
O_1^{{\cal N}}$ inherited from the never-vanishing factor ${\cal
F}(\sigma,\tau)=1/{\cal N}(\sigma,\tau)$.

In a way which generalizes the illustrative  Figure 2, each
hyperbolic singularity $H_k$ (defined by the equation
$R(\sigma,\tau)= (L-\ell)(k+1/2)\pi/\ell$ with $k = 0,1, \ldots$)
moves to the left with the growth of $\ell$ and $\lambda$. Once it
gets close to the $N-$th singularity oval $D_{N-1}$, it touches it
at a point with the coordinates $\sigma^{(N\!,\,k)}_{(in)} =0$ and
$\tau^{(N\!,\,k)}_{(in)} =(N-1/4)\pi$ at the critical value
$\lambda= 2\ell/\left ( L-\ell \right ) = (4k+2)/(4N-1) \equiv
\lambda^{(N\!,\,k)}_{(in)} $ of the shift.

With the further growth of $\lambda$ the intersection of the
hyperbola with the standing oval moves to the left and disappears,
curiously enough, at a certain pair of points with the
``last-contact" $|\sigma| =|\sigma^{(N\!,\,k)}_{(out)}|>0$ and
$\tau = \tau^{(N\!,\,k)}_{(out)}<(N-3/4)\pi$. The latter value
lies slightly below the oval's end. Let us skip here the proof of
this subtlety as not too relevant.

\subsection{A completion of the list of the nodal lines}

We are now prepared to detect {\em all} the nodal curves of ${\cal
D}[\sigma,\tau,R(\sigma,\tau)]$ and to determine their qualitative
$\ell-$dependence in all the interval of $\ell \in (0,L)$ and/or
of $\lambda=\lambda(\ell) \in (0,\infty)$. For the first
inspiration we return to Figure 2 where the oval of zeros $
O_1^{{\cal D}}\equiv V_1 $ cannot be interpreted as a mere small
perturbation of $O_1^{{\cal M}}$ in spite of the fact that the
singularity hyperbola $H_0$ still did not touch the singularity
oval $O_1^{{\cal N}} \equiv D_1$ since $\lambda =
0.275<\lambda^{(2\!,\,1)}_{(in)} = 2/7 \approx 0.286$.

Still, the much more important observation made in Figure~2
concerns the emergence of the new curve $W_0$ of the new zeros of
the function ${\cal D}$. At the chosen $\lambda$ this curve just
entered Figure~2 at its right side. Our next Figure~3 confirms
that the new nodal curve $W_0$ moves to the left and gets deformed
in a way reflecting the presence of a steep oval dip in
$Q(\sigma,\tau)$ below $\tau = 2\pi$. We choose $\lambda=0.355$
which is still safely smaller than the lower estimate $
(4k-2)/(4N-3) = 0.4$ of the singularity hyperbola's ``jumped-over"
parameter $ \lambda^{(2\!,\,1)}_{(out)} \approx 0.403$.

The ``next-step snapshot" of Figure~4 at $\lambda = 0.395$ shows
how the same dip deforms the shape of the oval $O_1^{{\cal
D}}\equiv V_1$ in the domain where the function of $R$ is small.
In the subsequent Figure~5 we finally see how the two curves of
the zeros merge while a topologically new situation is created and
sampled at $\lambda = 0.415>\lambda^{(2\!,\,1)}_{(out)} $.

We may summarize that for the growing $\lambda$ the motion of the
singular component $\tan 2\varrho(R)/R$ of our secular determinant
${\cal D}(\sigma,\tau)$ to the left gives a clear guide how to
keep the $\ell-$dependence of its zero lines under full control.
The emergence and the asymptotically hyperbolic shape of the new
(and, in fact, not quite expected) non-oval curves $W_m$ of zeros
follows immediately from the asymptotic smallness of the positive
component $Q(\sigma,\tau) \sim 1/\sigma^2$ of ${\cal D}
(\sigma,\tau) $ at the larger $|\sigma| \gg 1$.

Due to the reasonably elementary character of the function ${\cal
D} (\sigma,\tau) $ we are able to understand that the pattern
sampled by the Figures 2 - 5 is entirely universal. Always, step
by step, the nodal ovals $V_n$ as well as their asymptotically
hyperbolic nodal-line partners $W_m$ become deformed by the
existence of the dip in the numerator function ${\cal
M}(\sigma,\tau)$.

Of course, after the hyperbola of singularities $H_k$ as well as
its strongly deformed trailing nodal curve $W_k$ ``creep" over the
fixed singularity oval $D_j$ (as well as over its attached and
strongly deformed zero curve $V_j$), the smoother shapes of both
the nodal curves $W_k$ and $V_j$ are more or less recovered, and
only their ordering remains permanently reversed. In spite of the
apparent nonlinearity of the ``creeping-over" effects, their
details might again be analyzed algebraically, using an adapted
version of the moving-lattice method of section \ref{pohyb}.

The most important reward compensating an increase in complexity
of the latter recipe is that one becomes able to treat one of the
two roots of eq. (\ref{secudd}), say, as a ``non-perturbative"
solution at the small $\ell$. The most important example of its
role are the hyperbolic nodal curves $W_m$ which move to the right
in $\tau$ with the decrease of $\ell$ and which disappear in
infinity in the NSW limit of $\ell \to 0$.

\section{Energies}

\subsection{Graphical representation and classification}

On the background of the preceding material, what remains for us
to do is a combination of the above-described knowledge of the
nodal lines of ${\cal D}[\sigma,\tau,R(\sigma,\tau)]$ with the
coupling-dependence constraint $\sigma\times\tau = Z =
(L-\ell)^2g/2$. A sample of the intersections of this type (i.e.,
of a typical final solution) is offered in Figure 6 where
$\lambda=2.40$ is neither small nor large and where we choose
$Z=Z^{(a)} = 1.00$ and $Z=Z^{(b)} = 2.24$ (= the critical
``exceptional-point" value of ref. \cite{sqw}) for illustration.
The conclusions which are illustrated by this graph have a general
validity:

\begin{itemize}

 \item
We always have $\tau > \sigma>0$ which means that all the real
bound-state energies $E_n$ remain positive at $Z > 0$.

 \item
Some of the energies  remain real at {\em any} value of $Z> 0$.
They correspond to the intersections of the hyperbola
$\sigma=Z/\tau$ with the hyperbolic nodal lines $W_m$ and may be
called ``stable", $E=E_m^{(s)}$.

 \item
All the other energies $E=E_n^{(u)}$ correspond to the
intersections of the hyperbola $\sigma=Z/\tau$ with the nodal
ovals $V_k$. At a sufficiently small $Z$ the latter intersections
remain real (see the line (a) with $Z=1$ in Figure~6).

 \item
We may call the latter energies ``unstable" as they merge in pairs
and form complex conjugate doublets \cite{AM} beyond certain
``exceptional-point" \cite{Heiss} values of $\ell$ and $Z$
(illustration: the line (b) in Figure~6).

\end{itemize}

 \noindent
The decomposition of the spectrum into its stable and unstable
parts varies with $\ell$ or $\lambda = \ell/(L-\ell)$ in an
obvious manner. Hence, the stability pattern in the spectrum will
be entirely different at the small and large $\lambda$ since in
the former case the hyperbolic curves $W_n$ only generate the
high-lying energies and {\it vice versa}.

\subsection{Numerical
construction }

After all our previous detailed analysis of the qualitative
features of the spectrum the numerical determination of energies
becomes fully routine. Indeed, as long as we know
$\tau={Z}/{\sigma}$, the rule $\tau^2-\sigma^2=R^2$ leads
immediately to the definition of
 \be
 \sigma=\sigma(R) = \sqrt{\frac{2\,Z^2}{R^2+\sqrt{R^4+4\,Z^2}}}\,.
 \label{singlecross}
 \ee
In parallel to such an introduction of the closed function
$\sigma=\sigma(R)$ of $R$ we may return once more to the recipe
$\tau={Z}/{\sigma(R)}$ and re-read it as another explicit
definition of the second auxiliary function
$\tau(R)={Z}/{\sigma(R)}$ of $R$.

In such a setting, the purely numerical determination of the
bound-state energies is reduced to the search for the roots $R_n$
of eq. (\ref{secularia}), i.e., of the zeros of the secular
determinant
 \ben
 \hat{\cal D}(R) =
 \left [
 \sigma^2(R)\,\cosh 2\,\sigma (R)+ \tau^2(R)\,\cos 2\, \tau(R)
  \right ]\, {\sin 2\,\lambda \,R}
  \,+
 \een
 \be
  +R
  \,\left [
  \sigma (R)\,\sinh 2\,\sigma (R)
   +  \tau (R)\,\sin 2\,\tau (R)
   \right ]\,  {\cos 2\,\lambda \,R}
   \label{searia}
 \ee
converted now in the function of the single variable $R \sim
\sqrt{E}$. An illustration of such a search is given in Figure~7
at a fixed choice of $Z=2$. The quadruplet of the graphs of the
secular determinant $ \hat{\cal D}(R) ={\cal
D}[\sigma(R),\tau(R),R]$ is presented there at the four different
values 1.25, 1.35, 1.45 and 1.55 of $\lambda$ (indicated along the
vertical axis). In each of these graphs we magnified the vertical
units near $ \hat{\cal D}(R) \approx 0$ and compressed them to a
single point representing all the bigger values of $| \hat{\cal
D}(R) |\geq \varepsilon$. In this way the picture samples the
left-hand side of eq. (\ref{searia}) solely near its zeros. Our
magnification of the vertical dimension marks these zeros by the
virtually straight parts of the curve which are seen as
practically perpendicular to the horizontal axis.

The set of graphs in Figure 7 illustrates the $\lambda-$dependence
of the bound-state roots $R_n$. We see that a pair of the unstable
energies may merge and cease to be real after a fine-tuned growth
of $\lambda$. This illustrates the complexification of the
unstable energies which is {\em not} caused by the growth of $Z$
but rather by the growth of $\lambda$. At the first sight this
phenomenon looks like a paradox because we are now {\em weakening}
the non-Hermiticity in fact. Fortunately, this paradox is still
easily understood once we imagine (and check, say, in the spirit
of Figures 2 or 3) that the growth of $\lambda$ ``pushes" {\em
all} the zeros (including of course also the nodal oval in
question) to the left. Of course, this oval cannot get prolonged
in the $\sigma$ direction because the function ${\cal
M}(\sigma,\tau)$ itself grows too quickly with $\sigma$. This
implies that the two real intersections of the oval with the
hyperbola $\sigma=Z/\tau$ disappear because the latter curve grows
to the left.

In the light of an additional scaling in eq. (\ref{eneriad}) one
may only admire the subtlety of the phenomenon, the verification
of which very much profits from the exact solvability of the
model. An independent confirmation of the absence of any
contradictions may be also offered via a further simplification of
mathematics. This inspires us to pay particular attention to the
``most counterintuitive" limiting case where $L \to \infty$. Such
an analysis may be of an independent interest as it simulates,
very roughly, the shape of the most popular antisymmetric and
purely imaginary potential $V(x) \sim i\,x^3$ with real spectrum
\cite{DDT}. As long as this discussion already lies somewhat
beyond the scope the present text, it is moved to the Appendix~B.

\section{Conclusions \label{VI}}

After more than ten years of an intensive research many people now
seem to believe that we now better understand the key problems
related to the so called ${\cal PT}-$symmetric as well as to many
other similar non-Hermitian models or, in the more rigorous
terminology, to all the models where the metric remains
nontrivial, $\eta \neq I$ \cite{CZJ}. By the way, not all the
related results are new. For example, Scholz et al \cite{Geyer}
(inspired, presumably, by a few earlier mathematical as well as
physical publications) studied the similar $\eta \neq I$ models
more than ten years ago (!) and coined the name ``quasi-Hermitian"
for them.

Still, one cannot deny that during the last cca seven years, a new
and intensive excitement has been caused by the discoveries of the
reality of the spectra in many ${\cal PT}-$symmetric models. The
emphasis of the research has been shifted, typically, to the
explicit constructions of the charge ${\cal C}$ \cite{Cannata} or
to the more detailed analysis of what happens at the
``exceptional" points where the reality of the spectrum is being
lost \cite{Heiss,Patrick}. A few unusual features exhibited by our
present model seem to offer another welcome and clear intuitive
guidance in this area.

We found our results interesting since the merger and subsequent
spontaneous complexification of some ``twin" pairs $E^{(\pm
twin)}$ of the energies cannot be easily described within the
usual textbook models where the metric is ``trivial",
$\eta_{(trivial)} =I$. It is also in this context where
considerations based on our present model could lead to a deeper
insight in the underlying mechanisms and mathematics, not only
because our model is solvable but also because it proves able to
provide different ``twin-merging" patterns in the spectrum.
Indeed, by the choice of the shape parameter $\ell$ we may, up to
a large extent, prescribe {\em which} particular excitations (say,
in the low-lying spectrum) should remain robustly stable and which
ones should form the unstable, fragile ``twins" merging at some
sufficiently large couplings $g_{(critial)}$.

In the similar constructions and studies, one might feel hesitant
whether his/her models should be simpler or more realistic. We
believe that one should transfer the insight gained in the
solvable models (like in the present one) to all the more
realistic applications where just some approximate methods can be
used. In this sense we already mentioned a parallelism between the
role of the shift $\ell$ in our solvable model and of the exponent
$\mu$ in the power-law potentials with ${\cal PT}-$symmetry.

It is encouraging to see that a certain nontrivial enrichment of
the merging pattern has been detected, more or less in parallel,
within the class of the power-law forces~\cite{Patrick}. In this
comparison, our present model's merit lies in its exact
solvability. Definitely, it proves able to offer a comparably rich
pattern of the mergers of the levels.

This being said, the {\em key} phenomenological and
``model-building" specific merit of our present new version of the
${\cal PT}-$symmetric square-well model is still to be seen in the
``global" structure of its spectrum. There, one observes that the
``fragile" and the ``robust" levels seem to form the two sets
which may be moved with respect to each other as a whole. Thus,
the {\em whole} spectrum becomes ``almost completely robust" in
one extreme (which is ``almost Hermitian") and ``almost all
fragile" in another extreme which is, near $\ell \approx 0$,
``maximally non-Hermitian".

\subsection*{Acknowledgment}

Partially supported by GA AS in Prague, contract No. A 1048302.

\section*{Figure captions}

\subsection*{Figure 1.
A thin slice through the surface $Q(\sigma,\tau) = {\cal M}/{\cal
N}$. }

\subsection*{Figure 2.
A thin slice through the surface of the secular determinant ${\cal
D}(\sigma,\tau)$ at $\lambda = \ell/(L-\ell)=0.275$.
 }

\subsection*{Figure 3.
Same as Figure 2, $\lambda=0.355$. }

\subsection*{Figure 4.
Same as Figure 2, $\lambda=0.395$. }

\subsection*{Figure 5.
Same as Figure 2, $\lambda=0.415$. }

\subsection*{Figure 6.
Solutions at $Z^{(a)}=1.00$ and $Z^{(b)}=2.24$, intersections
marked by circles, $\lambda=2.4$. }

\subsection*{Figure 7.
Four re-scaled graphs of the function $ \hat{\cal D}(R)$. }

\subsection*{Figure 8. Graphical
solution of eq. (\ref{enep}) ($y = \omega_N/2$, $T=1$)}

\newpage

\section*{Appendix A: The method of moving lattice}

Secular eq. (\ref{secularia}) and its descendants contain quickly
oscillating trigonometric functions of arguments $2\tau$ and
$2\varrho$. In the spirit of ref. \cite{fragil} it makes sense to
re-parametrize both these variables according to the rules
 \ben
 {\tau}=\tau(N,t)
  = { \pi\, N + \pi\,t}\,,
 \ \ \ \ N = 0, 1, \ldots, \ \ \ \ \ t \in (0,1),
 \een
 \ben
 \varrho
 = \pi\,K+ \pi\,r\,,
 \ \ \ \ K = 0, 1, \ldots, \ \ \ \ \ r \in (0,1)\,
 \een
which separate their ``large" change (by an integer multiple of
the period $2\pi$ so that the trigonometric function itself
remains unchanged) from a ``small" change [within one period $(0,
2\pi)$]. Thus, once we define
 \ben
 \Psi=\sin\, 2\,\tau= \sin\, 2\,\pi\,t, \ \ \ \ \ \ \ \
 \Phi=\cos \,2\,\tau= \cos \,2\,\pi\,t\,,
 \label{firstpair}
 \een
 \ben
 \Xi=-\tan \,2\,\varrho = -\tan \,\pi\,r\,,
 \een
all our trigonometric functions in question become independent of
both the integer variables. Thus, once we decide to work, say, in
the $\sigma - R$ plane, we simply introduce a lattice ${\cal
L}_{t,r}$ of points with coordinates
 \ben
 {\sigma}=\sigma(N,t)
  = \frac{Z}{\tau(\sigma)} = \frac{Z}{ \pi\, N + \pi\,t}\,
 \een
and
 \ben
 R=
 \frac{L-\ell}{\ell}\,\varrho(R)
 = \frac{\pi\,(L-\ell)}{2\,\ell}\,(K+r)\,
 \een
where $ t \in (0,1)$ and $ r \in (0,1)$ are fixed while $ N = 0,
1, \ldots $ and $K = 0, 1, \ldots$ remain variable. Our secular
equation (\ref{secularia}) then becomes more easily analyzed at
the fixed $ t \in (0,1)$ and $ r \in (0,1)$ when it may be re-read
as a simplified mapping $\sigma \to R$ with
 \be
 R= {R}_{t,r}(\sigma)={\Xi}_r\,\times\, \frac{
 \sigma^4\,\cosh 2\,\sigma + Z^2\Phi_t
 }{\sigma^3\,\sinh 2\,\sigma
   + \sigma\,Z\,\Psi_t\,}\,,
   \label{securbcd}
 \ee
i.e., ${R}_{t,r} \approx {\Xi}_r\, | \sigma|$ at $|\,\sigma| \gg
1$ while
 \ben
  {R}_{t,r}\approx \frac{Z}{\sigma}\,\times\, \frac{
  \Phi_t\,{\Xi}_r
 }{\Psi_t}\,
   \label{secllcd}
 \een
at $ |\,\sigma| \ll 1$ etc. In the subsequent step, remembering
that the latter formulae hold on the lattice ${\cal L}(t,r)$ only,
we must let this lattice move with the variation of $t$ and/or
$r$. Within each box numbered by the pair $(N,K)$ of non-negative
integers we would be able to re-derive all the qualitative
geometric considerations of section \ref{tristar} in an
alternative, quantitative manner.

\newpage

\section*{Appendix B: Shallow well }

In the infinite-size limit $L \to \infty$ our model degenerates to
a purely imaginary square well with asymptotic boundary conditions
 \be
\psi(\pm \infty) = 0
 \label{bc}
 \ee
and with the ${\cal PT}$ symmetric matching conditions in the
origin,
 \be
  \psi(0)=1,
 \ \ \ \ \ \ \ \p_x \psi(0)=i\,G.
  \label{nor}
 \ee
This means that we have the general solution
  \be
 \psi(x) =
 \left \{
 \begin{array}{llc}
   \cos k\,x + B\,\sin k\, x,
    & x \in (0,\ell),& k^2 = E,\\
 (L + i\,N)\,\exp (-\sigma\,x)
  ,& x \in (\ell,\infty),&
\sigma^2=i\,T^2-k^2,
   \ea
   \right .
 \label{ansatzre}
  \ee
with $T=\sqrt{g}$ and with the the purely imaginary constant $B
=i\,G/k$.

\subsection*{B.1. Matching conditions at $x=\ell$}

Let us split $\sigma = p + i\,q$ in its real and imaginary part
with $p,q \geq 0$. This gives the rules $p^2+k^2=q^2$ and
$2pq=T^2$, easily re-parameterized in terms of a single variable $
{\alpha}$,
 \be
p=q\,\cos {\alpha}, \ \ \ \ \ k =q\,\sin {\alpha}, \ \ \ \ \ q
=\frac{T}{\sqrt{2\cos {\alpha}}}, \ \ \ \ \ \  {\alpha} \in (0,
\ell/2).
 \label{param}
 \ee
The standard matching at the point of discontinuity is immediate,
 \ben
 \cos k\ell + B\, \sin k \ell =
 (L + i\,N)\,\exp (-\sigma\,\ell),
 \een
 \ben
 -\sin k\ell + B\, \cos k \ell =
 -\frac{\sigma}{k}(L + i\,N)\,\exp (-\sigma\,\ell).
 \een
After we abbreviate $\sigma/k=-\tan \Omega\ell $, we get an
elementary complex condition of the matching of logarithmic
derivatives at $x=\ell$,
 \be
 G = -i\,k\,\tan(k+\Omega)\ell.
 \label{secular}
 \ee
Its real part defines our first unknown parameter, $G =
G({\alpha})$. Due to our normalization conventions, the imaginary
part of the right-hand-side expression must vanish, ${\rm
Re}[\tan(k+\Omega)\ell]=0$. An elementary re-arrangement of such
an equation acquires the form of an elementary quadratic algebraic
equation for $X=\tan k\ell$. Its two explicit solutions read
 \be
X_1 = \frac{p+q}{k}, \ \ \ \ \ \ \ \ \ \ X_2 = \frac{p-q}{k}
 \label{sice}
 \ee
or, after all the insertions,
 \be
 \tan \left [
{\frac{ \ell T \sin {\alpha}^{(+)}}{\sqrt{2 \cos {\alpha}^{(+)}}
 }} \right ]=
 {\rm tan} \left [ \frac{\ell-{\alpha}^{(+)}}{2} \right ],
 \label{sicep}
 \ee
 \be
 \tan \left [
{\frac{ \ell T \sin {\alpha}^{(-)}}{\sqrt{2 \cos {\alpha}^{(-)}}
 }} \right ]=
 \tan \left [- \frac{{\alpha}^{(-)}}{2} \right ].
 \label{sicem}
 \ee
These equations specify, in implicit manner, the two respective
infinite series of the appropriately bounded real roots
${\alpha}={\alpha}^{(\pm)}_n \in (0, \ell/2)$.

\subsection*{B.2. Energies}

For $ {\alpha} \in (0, \ell/2)$ the left-hand-side arguments in
eqs. (\ref{sicep}) and (\ref{sicem}) run from zero to infinity and
the functions oscillate infinitely many times from minus infinity
to plus infinity. In contrast, the limited variation of the
argument ${\alpha}$ makes both the right-hand side functions
monotonic, very smooth and bounded, $ {\rm tan}
[{(\ell-{\alpha}^{(+)})}/{2}] \in (1,\infty)$ and $ {\rm tan}[
{{\alpha}^{(-)}}/{2}] \in (0,1)$. This indicates that our roots
$k=k({\alpha}_n^{(\pm)})$ will all lie within well determined
intervals,
 \ben
 k_n^{(+)} \in
 \left ( n+\frac{1}{4},n+\frac{1}{2} \right ),
 \ \ \ \ \ \ \ \ n = 0, 1, \ldots,
 \een
 \ben
 k_m^{(-)} \in
 \left ( m+\frac{3}{4},m+{1} \right )
 \ \ \ \ \ \ \ \ m = 0, 1, \ldots.
 \een
An additional merit of parametrization (\ref{param}) lies in an
unambiguous removal of the tangens operators from both eqs.
(\ref{sicep}) and (\ref{sicem}). This gives
 \ben
 k_n^{(+)}= n+\frac{1}{2}-\frac{\omega_{n}^{(+)}}{4}
 , \ \ \ \ \ \ \ \ \
 k_m^{(-)}= m+{1}-\frac{\omega_m^{(-)}}{4},
 \ \ \ \ \ \ \ \ \ \ \omega_n^{(\pm)}=
 \frac{2{\alpha}_n^{(\pm)}}{\ell}\  \in \ (0, 1).
 \een
After a change of notation with $\omega_n^{(+)} =\omega_{2n}$ and
$\omega_n^{(-)}=\omega_{2n+1}$, we may finally combine the latter
two rules in the single secular equation
 \be
\sin \left (
 \frac{\ell}{2}\omega_N \right )=
 \frac{2N+2-\omega_N}{4T}\cdot \sqrt{2 \cos \left (
 \frac{\ell}{2}\omega_N \right )}
 \ \ \ \ \ \ \ \ N = 0, 1, \ldots,
 \ \ \ \ \ \ \ \ \
 \label{enep}
 \ee
In a graphical interpretation this equation represents an
intersection of a tangens-like curve with the infinite family of
parallel lines. This is illustrated in Figure~8. The equation
generates, therefore, an infinite number of real roots $\omega_N
\in (0,1)$ at all the non-negative integers $N = 0, 1, \ldots$.
The discrete spectrum is unbounded from above and remains
constrained by the inequalities
 \be
  \frac{(N+1/2)^2}{4}\ \leq \
E_N \  \leq \ \frac{ (N+1)^2}{4}\
 \label{ene}
 \ee
independently of the coupling $T$.

\subsection*{B.3. Wave functions }

Equation (\ref{secular}) in combination with eqs. (\ref{sicep})
and (\ref{sicem}) determines the real parameter
 \be
G= G^{(\pm)}=-\frac{k^2}{q\pm p}
 \label{ecu}
 \ee
responsible for the behaviour of the wave functions near the
origin [remember that $B =iG/k$ in eq. (\ref{ansatzre})]. For its
deeper analysis let us first introduce an auxiliary linear
function of $\omega$ and $N$,\
 \ben
  \sqrt{R(\omega_N,N)}=
 \frac{2N+2-\omega_N}{4T}\  \in \ \left (
 \frac{N+1/2}{2T}\ , \frac{N+1}{2T} \right )
 \een
and re-interpret our secular eq. (\ref{enep}) as an algebraic
quadratic equation with the unique positive solution,
 \be
 \cos \left (
 \frac{\ell}{2}\omega_N \right ) = \frac{1}{R(\omega_N,N)
 +\sqrt{R^2(\omega_N,N)+1}} \ .
 \label{secu}
 \ee
This is an amended implicit definition of the sequence $\omega_N$.
As long as the right hand side expression is very smooth and never
exceeds one, the latter formula re-verifies that the root
$\omega_N$ is always real and bounded as required.

In the weak coupling regime (i.e., in the domain of the large and
almost constant $R \gg 1$ with the small square-well height $T$ or
at the higher excitations), our new secular equation (\ref{secu})
gives a better picture of our bound-state parameters $\omega_N=
1-\eta_N$ which all lie very close to one. The estimate
 \ben
 \frac{\ell}{2}\,\eta_N = \arcsin \frac{1}{R+\sqrt{R^2+1}} \approx
 \frac{1}{2R} - \frac{5}{48\,R^3} + \ldots\
 \een
represents also a quickly convergent iterative algorithm for the
efficient numerical evaluation of the roots $\omega_N$. One can
conclude that in a way compatible with our {\it a priori}
expectations, the value of $p=p_N={\rm Re} \sigma \approx q/2R$ is
very close to zero and, as a consequence, the asymptotic decrease
of our wave functions remains slow. We have $q=q_N={\rm Im} \sigma
\approx k$ so that, asymptotically, our wave functions very much
resemble free waves $\exp( -i k x)$. In the light of eq.
(\ref{ecu}) we have also $\psi(x) \approx \exp( -i k x)$ near the
origin.

In the strong coupling regime (i.e., for very small $R$
representing, say, the low-lying excitations in a deep well with
$T \gg 1$) we get an alternative estimate
 \ben
 \frac{\ell}{4}\,\omega_N = \arcsin
 \sqrt{
 \frac{1}{2}
 \left [ R- \left (
 \sqrt{1+R^2}-1
 \right )
 \right ]
 }
  \approx
 \frac{1}{2}\,R-\frac{1}{4}R^2 + \ldots\  \ll  \frac{\ell}{4}.
 \een
In the extreme of $R \to 0$ the present spectrum of energies moves
towards (and precisely coincides with) the well known levels of
the infinitely deep Hermitian square well of the same width $I =
(-\ell,\ell)$. In this sense, the ``complex-rotation" transition
from the Hermitian well to its present non-Hermitian ${\cal PT}$
symmetric alternative proves amazingly smooth.

The wave functions exhibit the similar tendency. In the outer
region, they are proportional to $ \exp ( -px)$ and decay very
quickly since $p = {\cal O}(R^{-1/2})$. The parameter $G^{(\pm)}$
becomes strongly superscript-dependent,
 \ben
 G^{(+)} = -\frac{k^2}{q+p} ={\cal O}(R^{3/2}),
 \ \ \ \ \ \
 G^{(-)} = -(q+p)={\cal O}(R^{-1/2}).
 \een
This means that in the interior domain of $x \in (-\ell,\ell)$,
the wave functions with the superscript $^{(+)}$ and $^{(-)}$
become dominated by their spatially even and odd components $\cos
kx$ and $\sin kx$, respectively. In this sense, the superscript
mimics (or at least keeps the trace of) the quantum number of the
slightly broken spatial parity ${\cal P}$.


\newpage

\begin{thebibliography}{99}


\bibitem{DB}
D. Bessis and C. M. Bender, private communication

\bibitem{BB}
C. M. Bender and S. Boettcher, Phys. Rev. Lett. { 80} (1998) 5243.

\bibitem{Frank}
U. G\"{u}nther, F. Stefani and M. Znojil, math-ph/0501069, J.
Math. Phys., in print.

\bibitem{AM}
A. Mostafazadeh, J. Math. Phys. 43 (2002) 205 and 2814.

\bibitem{BBJerr}
A. Mostafazadeh,  A Critique of PT-Symmetric Quantum Mechanics
(arXiv: quant-ph/0310164, unpublished);

C. M. Bender, D. C. Brody and H. F. Jones, Phys. Rev. Lett. 92
(2004) 119902 (erratum).

\bibitem{Langer}
H. Langer and C. Tretter, Czechosl. J. Phys. 54 (2004) 1113.

\bibitem{MB}
A. Mostafazadeh and A. Batal, J. Phys. A: Math. Gen. 37 (2004)
11645.


\bibitem{AmerJ}
C. M. Bender, D. C. Brody and H. F. Jones, Am. J. Phys. 71 (2003)
1095;

A. Mostafazadeh,
Czech. J. Phys. 54 (2004) 1125;

M. Znojil, PT-symmetry, ghosts, supersymmetry and Klein-Gordon
equation (arXiv: hep-th/0408081), in ``Symmetry Methods in
Physics", Ed. \v{C}. Burd\'{\i}k et al on CD with ISBN
5-9530-0069-3 (JINR, Dubna, 2004).

\bibitem{sqw}
M. Znojil, Phys. Lett. A. 285 (2001) 7.

\bibitem{Geza}
M. Znojil and G. L\'{e}vai,  Mod. Phys. Lett. A 16 (2001) 2273.

\bibitem{Quesne}
B. Bagchi, S. Mallik and C. Quesne, Mod. Phys. Lett. A17 (2002)
1651.

\bibitem{pseudo}
P. A. M. Dirac, Proc. Roy. Soc. London A 180 (1942) 1;

W. Pauli, Rev. Mod. Phys. 15 (1943) 175;

M. Znojil, What is PT symmetry? preprint quant-ph/0103054v1,
unpublished;

M. Znojil, Conservation of pseudo-norm in PT symmetric quantum
mechanics, preprint math-ph/0104012, unpublished;

R. Kretschmer and L. Szymanowski, The Interpretation of
Quantum-Mechanical Models with Non-Hermitian Hamiltonians and Real
Spectra, preprint quant-ph/0105054, unpublished;

B. Bagchi, C. Quesne and M. Znojil,  Mod. Phys. Letters A 16
(2001) 2047;

A. Mostafazadeh, J. Math. Phys. 43 (2002) 3944 and 6343.

B. Bagchi and C. Quesne, Phys. Lett. A 300 (2002) 18;

A. Ram\'{\i}rez and B. Mielnik, Rev. Fis. Mex. 49S2 (2003) 130;

Z. Ahmed and S. R. Jain, Phys. Rev. E 67 (2003) 045106(R);

A. Mostafazadeh, Czech. J. Phys. 53 (2003) 1079;

F. Kleefeld,
 in ``Hadron Physics, Effective Theories of Low Energy
QCD", AIP Conf. Proc. 660 (2003) 325.

C. M. Bender, Czech. J. Phys. 54 (2004) 13.

\bibitem{fragil}
M. Znojil,
J. Math. Phys. 45 (2004) 4418.

\bibitem{JakubZ}
M. Znojil,
J. Phys. A: Math. Gen. 36 (2003) 7825;

V. Jakubsk\'{y}, Czech. J. Phys. 54 (2004) 67;

V. Jakubsk\'{y} and M. Znojil,
Czech. J. Phys. 54 (2004) 1101.

\bibitem{FeiZ}
S. Albeverio, S. M. Fei and P. Kurasov, Lett. Math. Phys. 59
(2002) 227;

M. Znojil,
J. Phys. A: Math. Gen. 36 (2003) 7639;

S. M. Fei, Czech. J. Phys. 54 (2004) 43.

\bibitem{BBjmp}
C. M. Bender, S. Boettcher and P. N. Meisinger, J. Math. Phys. 40
(1999) 2201.

\bibitem{experimental}
M. Znojil, Czech. J. Phys. 54 (2004) 151.

\bibitem{ptho}
M. Znojil, Phys. Lett. A 259 (1999) 220 and 264 (1999) 108;

G. L\'{e}vai and M. Znojil, J. Phys. A: Math. Gen.,  33 (2000)
7165;

B. Bagchi, S. Mallik and C. Quesne, Int. J. Mod. Phys. A 17 (2002)
51;

C. S. Jia, S. C. Li, Y. Li and L. T. Sun, Phys. Lett. A 300 (2002)
115;

A. Sinha, G. L\'{e}vai and P. Roy, Phys. Lett. A 322 (2004) 78.

\bibitem{BBJ}
C. M. Bender, D. C. Brody and H. F. Jones, Phys. Rev. Lett. 89
(2002) 270401.

\bibitem{Cannata}
C. M. Bender, Czech. J. Phys. 54 (2004) 1027;

H. F. Jones, Czech. J. Phys. 54 (2004) 1107;

E. Caliceti, F. Cannata, M. Znojil and A. Ventura,
Phys. Lett. A 335 (2005) 26;

B. Bagchi, A. Banerjee, E. Caliceti, F. Cannata, H. B. Geyer, C.
Quesne and M. Znojil,
hep-th/0412211, Int. J. Mod. Phys. A, in print.

\bibitem{Caliceti}
E. Caliceti, S. Graffi and M. Maioli, Commun. Math. Phys. 75
(1980) 51;

G. Alvarez, J. Phys.{ A: Math. Gen. 27} (1995) 4589;

E. Delabaere and D. T. Trinh, J. Phys.{ A: Math. Gen. 33} (2000)
8771;

G. A. Mezincescu, J. Phys.{ A: Math. Gen. 33} (2000) 4911;

C. R. Handy, Czech. J. Phys. 54 (2004) 57;

M. Bentaiba, S. A. Yahiaoui and L. Chetouani, Phys. Let. 231
(2004) 175.

\bibitem{DDT}
P. Dorey, C. Dunning and R. Tateo, J. Phys. A: Math. Gen. 34
(2001) 5679;

K. C. Shin, Commun. Math. Phys. 229 (2002) 543.

\bibitem{Herbst}
I. Herbst, Commun. Math. Phys. 64 (1979) 279.


\bibitem{Heiss}
C. Dembowski et al, Phys. Rev. Lett. 86 (2001) 787;

W. D. Heiss and H. L. Harney, Eur. Phys. J. D 17 (2001) 149;

U. G\"{u}nther and F. Stefani, J. Math. Phys. 44 (2003) 3097;

W. D. Heiss, Czech. J. Phys. 54 (2004) 1091.

\bibitem{CZJ}
cf. two special issues of Czechosl. J. Phys. 54 (2004), pp. 1 -
156 and 1005 - 1148.

\bibitem{Geyer}
F. G. Scholtz, H. B. Geyer and F. J. W. Hahne, Ann. Phys. (NY) 213
(1992) 74.

\bibitem{Patrick}
P. Dorey, A. Millican-Slater and R. Tateo,
J. Phys. A: Math. Gen. 38 (2005) 1305.

\end{thebibliography}
\end{document}